\documentclass[conference]{IEEEtran}
\IEEEoverridecommandlockouts
\usepackage{cite}
\usepackage{amsmath,amssymb,amsfonts}
\usepackage{algorithmic}
\usepackage{graphicx}
\usepackage{textcomp}
\usepackage{subcaption}
\usepackage{multirow}
\usepackage{soul}

\def\BibTeX{{\rm B\kern-.05em{\sc i\kern-.025em b}\kern-.08em
    T\kern-.1667em\lower.7ex\hbox{E}\kern-.125emX}}
\begin{document}

\title{Evaluation of IEEE 802.11ad for mmWave V2V Communications\\
\thanks{B. Coll-Perales and J. Gozalvez's work is supported by the Spanish Ministry of Economy, Industry, and Competitiveness, AEI,  and FEDER funds (TEC2017-88612-R, TEC2014-57146-R), and the Generalitat Valenciana (APOSTD/2016/049). 
}
}

\author{
\IEEEauthorblockN{Baldomero Coll-Perales\IEEEauthorrefmark{1}\IEEEauthorrefmark{2}, 
Marco Gruteser\IEEEauthorrefmark{2} and 
Javier Gozalvez\IEEEauthorrefmark{1}}
\IEEEauthorblockA{\textit{\IEEEauthorrefmark{1}UWICORE, Universidad Miguel Hernandez de Elche, Avda. de la Universidad sn, 03202, Elche (Spain)} \\
\textit{\IEEEauthorrefmark{2}WINLAB, Rutgers University, 671 Route 1 South, 08902, North Brunswick, N.J. (USA)}\\
\{bcoll, j.gozalvez\}@umh.es, 
\{bcoll, gruteser\}@winlab.rutgers.edu}
}

\maketitle

\begin{abstract}
Autonomous vehicles can construct a more accurate perception of their surrounding environment by exchanging rich sensor data with nearby vehicles. Such exchange can require larger bandwidths than currently provided by ITS-G5/DSRC and Cellular V2X. Millimeter wave (mmWave) communications can provide higher bandwidth and could complement current V2X standards. Recent studies have started investigating the potential of IEEE 802.11ad to support high bandwidth vehicular communications. This paper introduces the first performance evaluation of the IEEE 802.11ad MAC (Medium Access Control) and beamforming mechanism for mmWave V2V communications. The study highlights existing opportunities and shortcomings that should guide the development of mmWave communications for V2V communications.
\end{abstract}

\begin{IEEEkeywords}
mmWave; IEEE 802.11ad; autonomous vehicles; connected vehicles; MAC; V2V; vehicular communications
\end{IEEEkeywords}

\section{Introduction}
Autonomous driving is heavily dependent on the vehicles' sensing accuracy and capability. Several studies have highlighted that autonomous vehicles can improve their sensing capability through the exchange of sensor data with other nearby vehicles \cite{b1}, including RADAR (RAdio Detection And Ranging), LIDAR (Light Detection and Ranging), visual camera images, or raw GPS (Global Positioning System) data. Vehicles can fuse their sensor data with that received from other vehicles to create a more accurate view of the surrounding environment. This use case is referred to as cooperative or collective perception or sensing \cite{b2}. The bandwidth demand necessary to exchange all these sensor data will challenge existing V2X (Vehicle to Everything) standards (ITS-G5/DSRC and Cellular V2X). 

A candidate alternative to support the exchange of sensor data is mmWave vehicular communications \cite{b4}\cite{b5}. mmWave can provide high bandwidth communications using large arrays of antennas and highly-directive beams that compensate the propagation effects inherent in high frequency bands and introduce additional gains in terms of spatial sharing. Standardization and regulatory activities in the mmWave band (30-300 GHz) are already underway. The European Commission has allocated 1 GHz of spectrum at 63 GHz for Intelligent Transportation Systems (ITS) applications and has increased the maximum total transmission power limitations at these frequencies to enable longer distance links \cite{b6}. In the U.S, the Federal Communications Commission (FCC) allocated 7 GHz in the 57-64 GHz band for unlicensed mmWave communications \cite{b7}. IEEE 802.11ad is the Wireless Local Area Network (WLAN) standard for mmWave communications at 60 GHz \cite{b8}. IEEE 802.11ad introduces notable new features at the PHY and MAC layers for providing multi-Gbps data links. These features include directional multi-gigabit channel access and beamforming protocols.

\begin{figure*}[tb]
\centering
\includegraphics[width=4.6in]{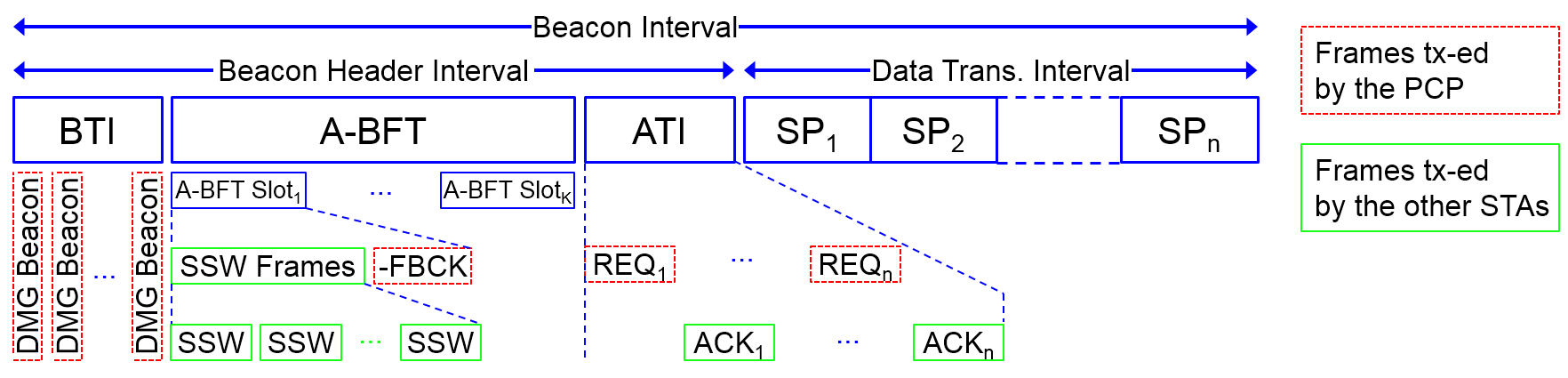}
\caption{IEEE 802.11ad channel access.}
\label{fig1}
\end{figure*}

Recent studies propose IEEE 802.11ad as a foundation for mmWave vehicular communications \cite{b4}\cite{b5}, just like IEEE 802.11a was the foundation for IEEE 802.11p. However, most of the existing studies focus on Vehicle-to-Infrastructure (V2I) and PHY layer aspects of mmWave communications. In this context, it is necessary to analyze the feasibility of mmWave V2V communications and to evaluate the operation and performance of the mmWave MAC protocols. This is particularly relevant given the highly dynamic vehicular network topology that can challenge the alignment between transmitter and receiver when highly-directional beams are utilized. This paper progresses the current state of the art by evaluating the use of IEEE 802.11ad to support Vehicle-to-Vehicle (V2V) mmWave communications. In particular, the paper focuses on the MAC of IEEE 802.11ad, and therefore complements the existing literature. The conducted evaluation highlights some inefficiencies of IEEE 802.11ad when utilized for mmWave vehicular communications that should be addressed in future work.

\section{Related Work}
The use of mmWave spectrum in automotive is not new since it is a frequency band utilized by automotive radars. In this context, Kumari et al. propose in \cite{b9} a joint vehicular communication-RADAR system using IEEE 802.11ad that reuses hardware and reduces cost and power consumption. The paper is focused on the radar functionality, and the authors show that IEEE 802.11ad can be used to detect vehicles with high accuracy and estimate the target vehicle's speed.  

Choi et al. justify in \cite{b4} the use of mmWave for vehicular communications on the need to exchange bandwidth-demanding sensor data for autonomous driving. The authors discuss three alternatives to support high data rates mmWave vehicular communications: 1) 5G mmWave cellular, 2) an adaptation of IEEE 802.11ad for vehicular communications, or 3) a dedicated new standard. The authors argue that IEEE 802.11ad could serve as a foundation for mmWave vehicular communications (see also \cite{b5}), but highlight the need to evolve the standard to suit the requirements and challenges characterizing vehicular communications. In this context, the authors propose using side-information (e.g. automotive sensors or DSRC) to reduce the beam alignment overhead in mmWave vehicular communications. The authors show that the overhead can be reduced by more than 90\% in V2I scenarios at 60GHz. Lochl et al. also seek in \cite{b10} to reduce the beam alignment overhead in mmWave V2I communications. In their case, the authors propose to use fix beams at vehicles and Road Side Units (RSU). The authors identify, using a software-defined-radio (SDR)-based prototype at 60GHz, the locations (e.g. on top of a bridge or inside a roundabout) that reduce the overhead as they maximize the time a vehicle is in the range of a RSU. 

Existing studies have focused on V2I communications and PHY layer aspects. mmWave vehicular communications can also be utilized for bandwidth-demanding V2V communications. In this case, it is important to design MAC protocols that will efficiently handle the challenges resulting from dynamic network topologies, blockage effects, and mmWave propagation impairments. This study progresses the current state of the art by providing, to the authors’ knowledge, the first analysis of mmWave V2V communications using IEEE 802.11ad. The study analyzes in detail the operation and possible inefficiencies of IEEE 802.11ad MAC processes when supporting V2V communications.

\section{IEEE 802.11ad Background}

IEEE 802.11ad is the amendment to the IEEE 802.11 standard for very high throughput (a.k.a. multi-gigabit communications) in the 60 GHz band. It operates in four non-overlapping channels of 2.16 GHz bandwidth each, but only channel 2 centered at 60.48 GHz is common in all regulatory domains. At the physical layer (PHY), IEEE 802.11ad supports three different modulation methods with a set of Modulation and Coding Schemes (MCSs): 1) control PHY (MCS 0); 2) single carrier (SC) and low power SC (MCS 1-12 and MCS 25-31, respectively); and 3) OFDM (MCS 13-24). The OFDM PHY provides the highest data rates. Data rates range from 693 Mbps with MCS13 to 6756.75 Mbps with MCS 24. The data rates for SC PHY range from 385 Mbps (MCS 1) to 4620 Mbps (MCS 12), and from 626 Mbps (MCS 25) to 2503 Mbps (MCS 31) for low power SC. The control PHY can provide data rates up to 27.5 Mbps, and it is mainly used when the link budget between the communicating stations is low (e.g. during the beamforming training phase).

The IEEE 802.11ad also introduces changes at the MAC layer to support Directional Multi-Gigabit (DMG) communications. The channel access is divided into Beacon Intervals (BIs). The BI is subdivided into access periods. Each access period has different access rules and serves a specific purpose. Fig.~\ref{fig1} shows an example of a BI comprising a Beacon Header Interval (BHI) and a Data Transmission Interval (DTI). The BHI is utilized to announce the network, and to conduct the beamforming training and the channel access scheduling. The beamforming helps compensate for increased signal propagation path loss at mmWave frequencies. Beamforming is achieved by sector sweeping that can last from tens to hundreds of milliseconds\cite{b15}. The DTI is the access period during which data frames are exchanged. The BHI includes the Beacon Transmission Interval (BTI), the Association Beamforming Training (A-BFT), and the Announcement Transmission Interval (ATI). Not all BIs must contain a BTI, and the presence of the A-BFT and ATI within the BHI is optional and signaled in the BTI. The DTI comprises Contention-Based Access Periods (CBAPs) and scheduled Service Periods (SPs). In CBAPs, stations contend for the channel access using the Enhanced Distributed Coordination Function (EDCF) mechanism. In SPs, the channel is reserved for communication between two dedicated stations. Any combination in the number and order of SPs and CBAPs in the DTI is possible.

IEEE 802.11ad defines a network type, referred to as Personal Basic Service Set (PBSS), in which stations can communicate directly with each other. PBSS is similar to the Independent Basic Service Set (IBSS) mode (a.k.a. ad-hoc) in IEEE 802.11, but with some important differences. Differently from IBSS, one station must assume the role of control point (PBSS Control Point, PCP) in a PBSS. The PCP station is the only station that can transmit beacons (DMG Beacons in IEEE 802.11ad) within a PBSS; in IBSS, all stations transmit beacons. The PCP station is also in charge of allocating/scheduling the CBAPs/SPs access periods to the rest of stations within the PBSS in the DTI.

\section{IEEE 802.11ad-based V2V Communications} \label{80211ad_for_V2V}
The IEEE 802.11ad standard sets the common framework for DMG communications, and defines mandatory mechanisms to ensure "basic" interoperability among stations \cite{b8}. However, it leaves open many implementation aspects, and defines optional mechanisms to improve performance and efficiency (e.g. clustering and relaying) \cite{b8}. This section presents the configuration of IEEE 802.11ad that has been implemented to analyze the feasibility of IEEE 802.11ad for mmWave V2V communications. 

This work considers that whenever a vehicle has data to transmit to its neighbor vehicles/stations that require the use of mmWave vehicular communications, it adopts the role of a PCP station and forms a PBSS. This work does not focus on any particular application or use case.  

The topology of vehicular networks is highly dynamic. As a result, it is necessary to frequently re-evaluate the beamforming to ensure high link budget communications. The beamforming is performed during the BTI and A-BFT access periods\footnote{Beamforming refinement could also be performed during the DTI access period, although this option has not been implemented in this work.}. To this aim, we assume that it is beneficial for mmWave V2V communications that the BI interval always contains the BTI and A-BFT --the ATI and DTI access periods are also included in the BI to conduct the channel access scheduling. 

During the BTI and A-BFT access periods, DMG beacons and SSW (Sector Sweep) frames are transmitted sequentially across each antenna sector\footnote{\label{note1}In this study, the first antenna sector is selected randomly, and subsequent transmissions follow a clockwise sequence.} (see Fig.~\ref{fig1}). DMG beacons are transmitted by the PCP station and SSW frames by the stations (referred to as responders while they participate in the beamforming process \cite{b8}) that receive any DMG beacon. DMG beacons and SSW frames include, among other fields, the MAC address of the transmitter, the antenna sector identification (ID), and a countdown (CDOWN). The antenna sector ID uniquely identifies the transmit antenna sector employed by the transmitter when transmitting each DMG beacon/SSW frame. The CDOWN field describes how many DMG beacons/SSW frames transmissions are still pending until the end of the BTI/A-BFT access period. The responder also includes in the SSW frame the antenna sector ID included in the DMG beacon frame received from the PCP station. Upon reception of the SSW frame, the PCP station identifies the antenna sector ID it has to use in the PCP-responder link, and the antenna sector ID that the responder should use in the responder-PCP link. This information is included in the sector sweep feedback (‘-FBCK’ in Fig.~\ref{fig1}) frame transmitted from the PCP station to the responder. Upon reception of the SSW-FBCK, the beamforming is completed. The following implementation decisions have been made in this study: 
\begin{itemize}
  \item The responder stations that have received (at least) one DMG beacon select randomly one of the A-BFT slots. The number of available A-BFT slots in the BI is included in the DMG beacon.
  \item If the PCP station receives more than one SSW frame from different responder stations during the same A-BFT slot, it sends a SSW feedback frame to each of them.
  \item During the handshaking in the BTI and A-BFT access periods, the receiving stations are set in (quasi-) omnidirectional mode. This mode is selected since the stations do not know the location of the transmitter (i.e., the direction from which the frame will arrive). This may result in low link budgets and therefore the PHY control mode (MCS0) is used for the transmission of the DMG beacons, SSW and SSW-FBCK frames.
\end{itemize}

During the ATI access period, request (REQ in Fig.~\ref{fig1}) and response (ACK in Fig.~\ref{fig1}) frames are exchanged between the PCP station and the stations that have successfully completed the A-BFT handshaking. The ATI access period takes place with beam-trained stations, and hence higher order MCS can be utilized. This work considers the use of OFDM PHY with MCS13 for the ATI handshaking. The PCP station uses directional links with the antenna sector pointing towards the station addressed in the REQ frame at each point in time. The stations keep using the antenna sector that points towards the PCP station during all the ATI access period since they do not know when are going to be contacted. When a station receives a REQ frame from the PCP station, it sends back an ACK frame. The REQ frame includes the SP access period the PCP station has selected to communicate with the station\footnote{This work does not consider CBAPs access periods.}. This work assumes that the number of REQ frames sent by the PCP station matches the number of SP access periods available in the DTI. In this study, if the number of stations that have successfully completed the A-BFT handshaking is higher than the available SP access periods, the PCP station selects randomly a subset of stations to which it will transmit REQ frames. The PCP station schedules each SP transmission with a specified start time and with a fixed duration. The SP access periods are allocated sequentially (i.e. $REQ_{1}$ allocates $SP_{1}$ and so on). For the allocated SP access period, the access to the channel within the PBSS network is granted to the PCP and the scheduled station (EDCF is used too to mitigate interference). During the DTI access period, the PCP station transmits data frames to the scheduled stations in the allocated SPs. The PCP and scheduled stations use the identified antenna sectors during the SP access period.

\section{Evaluation Scenario}
This study is conducted using the ns-3.26 simulation platform, and leveraging the IEEE 802.11ad implementation in \cite{b12}. Additional features (e.g. blockage detection and path loss model for inter-vehicle communications at PHY, and the MAC configuration described in Section \ref{80211ad_for_V2V}) necessary to simulate IEEE 802.11ad-based V2V communications have been added. 

The simulation scenario emulates a 16m x 80m highway section with 4 lanes. For each simulation snapshot, 10 cars are randomly deployed in the different lanes. Each snapshot simulates 2 seconds of mmWave communications. Over two hundred snapshots are simulated for each configuration of simulation parameters to ensure the statistical accuracy of the results. In each snapshot, each car has a \{10, 20, 30, 40\}\% probability of being selected as mmWave transmitter and hence of taking the role of PCP station. Each PCP needs to transmit 600 packets of 1600 bytes each during the SP access period. The  packet size has been selected following the 'collective perception of environment' use case discussed for eV2X under 3GPP Release 15 \cite{b13}. This work considers that the traffic generated by the PCP station is of interest to any car/station nearby the PCP station. This also includes other PCP stations.  

ns-3 computes the signal power level at the receiver $P_{RX}$ in dBm as $P_{RX}(dBm)=G_{TX}(dB)+G_{RX}(dB)+P_{TX}(dBm)-PL(dB)$, 
where $G_{TX}$ and $G_{RX}$ are the transmitter's and receiver's antenna gain, respectively, $P_{TX}$ is the transmission power and $PL$ is the propagation loss. This work considers an analog beamforming architecture limited to single-stream transmissions (digital and hybrid beamforming architectures are yet difficult to realize \cite{b5}). 
The antenna is approximated using a sectored antenna model. This study uses the $<<$ns3::Directional60GhzAntenna$>>$ antenna model available in \cite{b12} that equally divides the horizontal plane into a number of virtual sectors. The antenna radiation pattern is composed of a main lobe in the selected antenna sector, and of side lobes in the rest of antenna sectors. The radiation pattern of the main lobe has a maximum gain in the center angle of the sector. As a result, the $G_{TX}$ and $G_{RX}$ antenna gains depend on the geometric angle between the transmitter and the receiver and their selected antenna sectors.

\begin{table}[!t]
\caption{Simulation Parameters}
\begin{center}
\begin{tabular}{|c|c|}
\hline
\textbf{Parameter}&\textbf{Value} \\
\hline
\multicolumn{2}{|c|}{\textbf{\textit{PHY}}} \\
\hline
Antenna sectors & 14 \\
\hline
$P_{TX}$(dBm) & 10 \\
\hline
\multicolumn{2}{|c|}{\textbf{\textit{Path loss model}}} \\
\hline
\multirow{4}{*}{A; C} & 1.77; 70 (LOS), \\
                   & 1.71; 78.6 (1 vehicle), \\
                   & 0.635; 115 (2 vehicles), \\
                   & 0.362; 126 (3 vehicles) \\
\hline
\multicolumn{2}{|c|}{\textbf{\textit{IEEE 802.11ad access periods}}} \\
\hline
BTI, A-BFT, ATI [TUs$^{\mathrm{a}}$] & 5, [15, 20, 25, 30, 35, 40], 5 \\
\hline
$SP$ (ms) & 50 \\
\hline
$TU$ ($\mu$s) & 1024 \\
\hline
\end{tabular}
\label{tab1}
\end{center}
\end{table}

The propagation loss is modeled using the empirical path loss model presented in \cite{b14} for inter-vehicle communications at 60GHz. In \cite{b14}, the authors measure the path loss when there is Line-Of-Sight (LOS) between the transmitter and the receiver, and when there is one, two and three blocking vehicles between the transmitter and the receiver. In \cite{b14}, the authors use horn antennas mounted at the bumper level. To the authors’ knowledge, this model is the most suitable one to date for mmWave vehicular communications. \cite{b14} models $PL$ in dB as 
$PL(dB) = A\cdot 10 \cdot log_{10}(d)+C+15\cdot d /1000,$ 
where $d$ is the distance between the transmitter and the receiver (in meters), and $A$ and $C$ are constants that depend on the visibility conditions and number of blocking vehicles between the transmitter and the receiver (see TABLE ~\ref{tab1}).

The $PL$ model in \cite{b14} requires detecting the number of blocking vehicles between a transmitter and receiver. To this aim, this study has implemented a blockage detection function in ns3. The implemented function is based on an abstract representation of the real deployment of cars that is made considering the vehicles dimensions (2D; 5m length x 2m width) and location (from $<<$ns3::MobilityModel$>>$). This function identifies the number of blocking vehicles between the transmitter and the receiver by tracing a ray between them and counting the number of obstacles (i.e. rectangles defined by the abstract representation of the cars) the ray is passing through. 



\section{Results}

\begin{figure}[!t]
\centering
\includegraphics[width=3.3in]{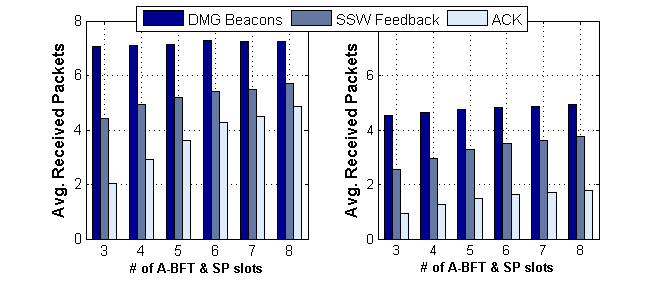}
(a) Prob. PCP station = 10\%    \hspace*{10pt}      (b) Prob. PCP station = 40\%        
\caption{Avg. number of received packets during the BHI access period as a function of the number of A-BFT and SP slots.}
\label{fig3}
\end{figure}

A preliminary study, considering ideal MAC and beamforming, has shown that each PCP station (mmWave transmitter) could contact on average 7.1 neighbor vehicles/stations. This performance is used as a benchmark for the results reported in Fig.~\ref{fig3}. 
Fig.~\ref{fig3} shows the average number of received packets during the IEEE 802.11ad BHI access periods. The figure depicts the average number of DMG beacons, SSW feedbacks and ACKs. These frames have been selected since they allow to keep track of the number of stations that successfully complete each handshaking in the BTI, A-BFT and ATI, respectively. The results in Fig.~\ref{fig3} are depicted for different numbers of A-BFT and SP slots in the BI, and for scenarios where the probability of a vehicle acting as a PCP station is 10\% (Fig.~\ref{fig3}.a) and 40\% (Fig.~\ref{fig3}.b). The number of received DMG beacons provides an indication of the number of neighboring vehicles to which a PCP station could transmit its data. For example, Fig.~\ref{fig3}.a shows that, under the simulated scenario, a PCP station could potentially communicate with up to 7 vehicles on average (i.e. similar to the benchmark results). The vehicles (responder stations) that receive the DMG beacons select randomly an A-BFT slot to complete the beamforming training with the PCP station. Responder stations autonomously select their A-BFT slot, so it is possible that several stations select the same A-BFT slot. The responders transmit then SSW frames through all their antenna sectors. A collision between SSW frames will occur at the PCP station if several responders transmit a SSW frame in the direction of the PCP station at the same time. Other effects (radio propagation, interference, antennas misalignment, etc.) can also cause an incorrect reception of SSW frames at the PCP station. The PCP station then sends back a SSW feedback frame to those responders from which it has correctly received a SSW frame. If the PCP station receives more than one SSW frame from different responder stations during the same A-BFT slot, it sends a SSW feedback frame to each of them. Fig.~\ref{fig3} also depicts the SSW feedback frames received by those stations. The results in Fig.~\ref{fig3} show that the number of responder stations receiving SSW feedback frames increases with the number of A-BFT slots in the BI. This is the case because increasing the number of A-BFT slots reduces the probability that more than one responder station will select the same A-BFT slot and so the interference and potential collisions during the A-BFT slots. Once the A-BFT access period is completed, the PCP station sends the REQ frames to schedule transmissions in the SP access periods in the DTI. The addressed stations send back an ACK frame to the PCP station if they receive the REQ frame. The number of REQ frames sent by the PCP station is limited by the available SP access periods or the received SSW feedback frames. In the scenario corresponding to Fig.~\ref{fig3}.a, the PCP station sends only 3 and 4 REQ frames when the BI is configured with 3 and 4 A-BFT and SP slots, respectively, even if it received on average a higher number of SSW feedback frames (on average 4.5 and 5, respectively). Therefore, the PCP station discards a subset of stations in the ATI access period. For higher number of A-BFT and SP slots in Fig.~\ref{fig3}.a, the number of REQ frames sent by the PCP station is limited by the number of received SSW feedback frames; this number is lower than the number of available SP slots. The results reported in Fig.~\ref{fig3}.a show that the number of received ACK frames increases with the available number of A-BFT and SP slots. This suggests that increasing the number of A-BFT and SP slots would  increase the number of stations scheduled in the DTI access period, and hence the network performance. It should be noted that the achieved performance (i.e. stations scheduled in the DTI) is lower than the benchmark results which showed 7.1 neighbor vehicles/stations can be contacted on average. Fig.~\ref{fig3}.b shows similar trends to those observed in Fig.~\ref{fig3}.a but with lower values. This reduction is caused by the higher number of mmWave transmitters or PCP stations in the scenario. This study's IEEE 802.11ad implementation allows PCP stations to communicate with each other. However, the lack of coordination among them results in multiple conflicts, including interference and collisions during the beamforming (BTI and A-BFT access periods) and scheduling (ATI access period) management. Such conflicts are at the origin of the differences between Fig.~\ref{fig3}.b and Fig.~\ref{fig3}.a. 

\begin{figure}[!t]
\centering
\includegraphics[width=3.3in]{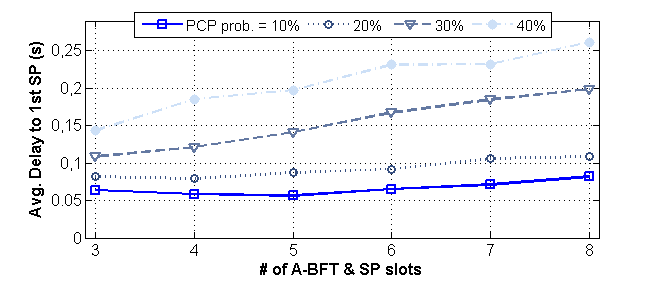}
\caption{Time elapsed to the start of the first SP access period.}
\label{fig4}
\end{figure}

In order to evaluate the IEEE 802.11ad's overhead and delay, Fig.~\ref{fig4} shows the average delay or time elapsed between the generation of the data packets at the PCP station and the start of the first allocated SP access period. The depicted results show that the delay increases with the number of PCP stations in the scenario and with the number of A-BFT and SP slots in a BI. It should be noted that the BTI, A-BFT and ATI access periods account for a fix share of the delay represented in Fig.~\ref{fig4}. In particular, they account for \{25.6, 30.72, 35.84, 40.96, 46.08, 51.20\} ms delay (overhead) when the BI is configured with \{3, 4, 5, 6, 7, 8\} A-BFT slots, respectively. In this context, and in contrast to what happen in Fig.~\ref{fig3}, increasing the number of A-BFT slots has a negative impact on the delay. The lack of coordination among PCP stations can result in that REQ and/or ACK frames are not correctly received by the addressed station. This provokes that certain SP access periods are not allocated to any station, and the delay represented in Fig.~\ref{fig4} can increase\footnote{It is important to remember that the SP access periods are allocated sequentially starting from the first one.}. The SP access periods that have not been correctly allocated to any station, and that occur before the first SP access period used by the PCP to transmit data packets, are also taken into account in the delay depicted in Fig.~\ref{fig4}. The SP access periods are 50ms long in this study. 

\begin{figure}[!t]
\centering
\includegraphics[width=3.3in]{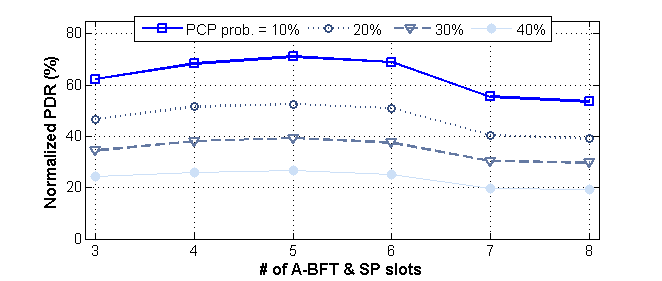}
\caption{Normalized packet deliver ratio.}
\vspace{-10pt}
\label{fig5}
\end{figure}

Fig.~\ref{fig3} and Fig.~\ref{fig4} highlight the existing trade-off between mmWave V2V performance (stations scheduled in the DTI access period, in Fig.~\ref{fig3}) and overhead/delay (Fig.~\ref{fig4}) when modifying the configuration of the BI. Fig.~\ref{fig5} seeks to find what would be a compromise configuration of the BI for the scenario under study. To this aim, Fig.~\ref{fig5} illustrates the normalized PDR (Packet Delivery Ratio) as a function of the number of A-BFT and SP slots. The normalized PDR is measured as the ratio between the data packets correctly received during the DTI access period, and the data packets that would be transmitted in the DTI access period if all SP access periods were correctly allocated. The low normalized PDR rates shown in Fig.~\ref{fig5} are caused by the fact that many SP access periods are not allocated to any station (see Fig.~\ref{fig3}). In the allocated SP access periods, the ratio of packets correctly received is above 90\% for all settings. Fig.~\ref{fig5} shows that the highest normalized PDR is achieved in this scenario when the BI is configured with 5 A-BFT and SP slots. However, it is important to highlight that the most adequate BI configuration will depend on context conditions (e.g. density of vehicles, mmWave transmission range, PHY mode, etc.), and on the application requirements.

\begin{figure}[!t]
\centering
\includegraphics[width=3.3in]{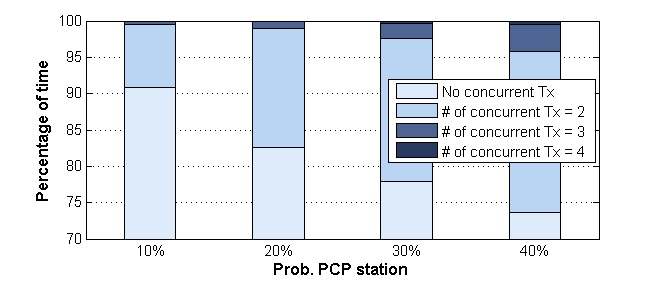}
\caption{Percentage of time there are concurrent transmissions in the scenario --the colors differentiate the number of concurrent transmissions (A-BFT \& SP slots = 4; similar trends are shown for the rest of BI configurations).}
\label{fig6}
\end{figure}

Finally, Fig.~\ref{fig6} analyzes the spatial sharing of IEEE 802.11ad-based V2V communications. Fig.~\ref{fig6} shows the percentage of time there are no concurrent transmissions and when there are \{2, 3 or 4\} concurrent transmissions for scenarios where the probability of a vehicle acting as a PCP station (mmWave transmitter) is \{10, 20, 30, 40\}\%. Based on the results shown in Fig.~\ref{fig3}, each PCP station has the potential to communicate with at least 4 vehicles (see ‘DMG beacons’ in Fig.~\ref{fig3}). The results in Fig.~\ref{fig6} are then reported for the case the BI is configured with 4 A-BFT and SP slots. In this case, each PCP station could potentially allocate one station to each SP access period. The results reported in Fig.~\ref{fig6} show that as the number of PCP stations in the scenario increases, the percentage of time there are concurrent transmissions also increases. However, Fig.~\ref{fig6} shows that even when on average 40\% of the vehicles are PCP stations (potentially 4 concurrent transmissions should happen the whole time), 73\% of the time there are no concurrent transmissions. The lack of coordination among PCP stations results in that certain SP access periods are not allocated to any station and hence the number of concurrent transmissions reduces. The reported results highlight that to fully exploit IEEE 802.11ad-based vehicular communications a mechanism that enables a coordination among PCP stations is necessary.

\section{Discussion and Conclusions}
This paper has evaluated in ns-3 the performance of IEEE 802.11ad for mmWave V2V communications. The study has focused on the MAC operation and beamforming processes while taking into account relevant mmWave propagation effects and models. An implementation of the IEEE 802.11ad BI tailored for vehicular networks has been presented. The study has shown that beamforming and scheduling management in mmWave V2V communications result in a high overhead (up to 250ms). 
This study has also highlighted that there is a performance-overhead trade-off in the BI configuration that might benefit from the use of context-aware policies. This study has also found that the lack of coordination among PCP stations (mmWave transmitters) can significantly degrade the network performance. 

The IEEE 802.11ad standard includes several optional mechanisms that future work could study; including the enhancements introduced in IEEE 802.11ay. First, is a clustering protocol to facilitate the communication among PCP stations. The clustering protocol can improve the spatial sharing and mitigate interference by scheduling transmissions from different PBSS networks in non-overlapping time periods. The implementation of this protocol requires a Synchronization PCP (S-PCP) station in the range of all interfering PCP stations to coordinate their transmission. 
IEEE 802.11ad also includes a mechanism to improve spatial sharing within a PBSS. The mechanism requires stations to measure and share the signal level detected during the SP access period being assessed. Based on these measurements, the PCP station decides whether to allocate a concurrent transmission in the next BI. The delay in making the decision appears ill-suited for the dynamic and varying network topologies of vehicular networks since conditions can be different in the next BI.

mmWave vehicular communications have the potential to support bandwidth-demanding connected and autonomous vehicle use cases. However, this study has highlighted that changes are needed to the current IEEE 802.11ad standard to efficiently support mmWave V2V communications.


\end{document}